\documentstyle[11pt]{article}
\normalbaselines
\begin{document}
\bibliographystyle{unsrt}

\newtheorem{theorem}{Theorem}
\newtheorem{lemma}{Lemma}
\newtheorem{proposition}{Proposition}

\def\bea*{\begin{eqnarray*}}
\def\eea*{\end{eqnarray*}}
\def\ba{\begin{array}}
\def\ea{\end{array}}
\count1=1
\def\be{\ifnum \count1=0 $$ \else \begin{equation}\fi}
\def\ee{\ifnum\count1=0 $$ \else \end{equation}\fi}
\def\ele(#1){\ifnum\count1=0 \eqno({\bf #1}) $$ \else \label{#1}\end{equation}\fi}
\def\req(#1){\ifnum\count1=0 {\bf #1}\else \ref{#1}\fi}
\def\bea(#1){\ifnum \count1=0   $$ \begin{array}{#1}
\else \begin{equation} \begin{array}{#1} \fi}
\def\eea{\ifnum \count1=0 \end{array} $$
\else  \end{array}\end{equation}\fi}
\def\elea(#1){\ifnum \count1=0 \end{array}\label{#1}\eqno({\bf #1}) $$
\else\end{array}\label{#1}\end{equation}\fi}
\def\cit(#1){
\ifnum\count1=0 {\bf #1} \cite{#1} \else 
\cite{#1}\fi}
\def\bibit(#1){\ifnum\count1=0 \bibitem{#1} [#1    ] \else \bibitem{#1}\fi}
\def\ds{\displaystyle}
\def\hb{\hfill\break}
\def\comment#1{\hb {***** {\em #1} *****}\hb }

\newcommand{\TZ}{\hbox{\bf T}}
\newcommand{\MZ}{\hbox{\bf M}}
\newcommand{\ZZ}{\hbox{\bf Z}}
\newcommand{\NZ}{\hbox{\bf N}}
\newcommand{\RZ}{\hbox{\bf R}}
\newcommand{\CZ}{\,\hbox{\bf C}}
\newcommand{\PZ}{\hbox{\bf P}}
\newcommand{\QZ}{\hbox{\rm eight}}
\newcommand{\HZ}{\hbox{\bf H}}
\newcommand{\EZ}{\hbox{\bf E}}
\newcommand{\GZ}{\,\hbox{\bf G}}

\font\germ=eufm10
\def\goth#1{\hbox{\germ #1}}
\vbox{\vspace{38mm}}

\begin{center}
{\LARGE \bf Bethe Ansatz and Symmetry in Superintegrable Chiral Potts Model and 
Root-of-unity Six-vertex Model\footnote{Talk presented at the 23rd International
Conference ``Differential Geometric Methods in Theoretical Physics (DGMTP)'' at Nankai Institute, Tianjin, China, Aug. 20-26, 2005.}}\\[10 mm] 
Shi-shyr Roan\footnote{\uppercase{T}his work is supported in part by National Science Council of Taiwan under Grant No NSC 94-2115-M-001-013.} \\
{\it Institute of Mathematics \\
Academia Sinica \\  Taipei , Taiwan \\
(email: maroan@gate.sinica.edu.tw ) } \\[25mm]
\end{center}

\begin{abstract}
We examine the Onsager algebra symmetry of $\tau^{(j)}$-matrices in the superintegrable chiral Potts model. The comparison of Onsager algebra symmetry of the chiral Potts model with the $sl_2$-loop algebra symmetry of six-vertex model at roots of unity is made from the aspect of functional relations using the $Q$-operator and fusion matrices. The discussion of Bethe ansatz for both models  is conducted in a uniform manner through the evaluation parameters of their symmetry algebras.    
\end{abstract}
\par \vspace{5mm} \noindent
{\it 2003 PACS}:  05.50.+q, 02.20.Uw, 75.10Jm \par \noindent

\section{Introduction}
The symmetry of quantum spin chains and the related lattice models has recently attracted certain attention due to their close connection with diverse areas of physics as well as mathematics.  However, up to the present stage, only limited knowledge is available about the symmetry of lattice vertex models, and few exact results are obtained in this area. Even the $sl_2$-loop algebra symmetry of the six-vertex model at roots of unity, found in \cite{DFM}, has not been fully understood till now, given that much accomplishment has been made on the study of evaluation parameters related to the symmetry algebra representation. Some conjectures supported by numerical evidence remain to be answered, (see \cite{De05} \cite{FM01} and references therein). Though the understanding of the symmetry of eight-vertex model in \cite{FM02} \cite{FM04} is still rudimentary in the present stage, 
the conjectural functional-relation-analogy discovered in the study on the eight-vertex model and the $N$-state chiral Potts model (CMP)\footnote{In this paper, the discussion of the chiral Potts model will be confined only to the full homogeneous lattice by taking $p=p'$ in \cite{BBP}.}  in \cite{BBP} did lead to exact results about the Onsager algebra symmetry of $\tau^{(j)}$-models in the superintegrable CPM \cite{R04} \cite{R05}. In the study of CPM, 
the lack difference-property for the rapidities is considered as the characteristic nature which distinguishes CPM from other known solvable lattice models. Nevertheless, progress made on the transfer matrix of CPM for the past fifteen years, culminating in the recent Baxter's proof of the order parameter \cite{B05}, has provided the sufficient knowledge for the  understanding of the detailed nature about the symmetry of superintegrable CPM. By this, the study of CPM could suggest a promising method to help the symmetry study about the six-vertex model at roots of unity as the limit case of eight-vertex model from the approach of functional relations, a scheme proposed in \cite{FM04}. In this paper, we examine the similarity of the symmetry structure of two lattice models: the superintegrable CPM and the six-vertex model at roots of unity. The symmetry of superintegrable CPM is described by the Onsager algebra, obtained in \cite{R04} \cite{R05}, with a short explanation in Sec. 2. The six-vertex model at roots of unity possesses the $sl_2$-loop algebra symmetry by the works in \cite{DFM} \cite{De05} \cite{FM01}. We present a discussion of symmetry of six-vertex model, parallel to the theory of CPM, through Bethe roots and evaluation parameters in Sec. 3, and give some concluding remarks in Sec. 4. 

{\bf Notations.}
In this paper, $\ZZ, \CZ$ denote 
the ring of integers, complex numbers
respectively, $N$ is a positive integer $\geq 2$, $\ZZ_N=
\ZZ/N\ZZ$,  and ${\rm i} = \sqrt{-1}$, $\omega= e^{\frac{2 \pi i}{N}}$. 
Denote by $X, Z$  the Weyl 
operators of the vector space $\CZ^N $, defined by  $X |n \rangle = | n +1 \rangle$, $ Z |n \rangle = \omega^n |n \rangle $ for $n \in \ZZ_N$, and $\stackrel{n}{\otimes} \CZ^N$ is the $n$-tensor product of $\CZ^N$.

\section{The $N$-state Chiral Potts Model}
\subsection{Rapidity and functional Relation of chiral Potts model}\label{subsec:fun}
In the study of CPM as a descendent of the six-vertex model, Bazhanov and Stroganov obtained the following 3-parameter family of Yang-Baxter
solutions for the inhomogeneous R-matrix of six-vertex model,
$$
R(t ) = \left( \begin{array}{cccc}
        t \omega - 1  & 0 & 0 & 0 \\
        0 &t-1 & \omega  - 1 &  0 \\ 
        0 & t(\omega  - 1) &\omega( t-1) & 0 \\
     0 & 0 &0 & t \omega - 1    
\end{array} \right) \ ,
$$
with the $\CZ^N$-operator entries parametrized by a four-vector ratio $p=[a, b, c, d]$ \cite{BazS, R04}:
\bea(l)
 b^2 G_p (t)  = \left( \begin{array}{cc}
       b^2 - t d^2 X & ( bc - \omega a d X )  Z   \\
       -t (b c - a d X ) Z^{-1} & - t c^2  + \omega  a^2  X  
\end{array} \right)  \ , \ \ \
t \in \CZ \ ,
\elea(G)
which satisfy the following Yang-Baxter equation: 
$$
R(t/t') (G_p (t) \bigotimes_{aux}1) ( 1
\bigotimes_{aux} G_p(t')) = (1
\bigotimes_{aux} G_p(t'))(G_p (t)
\bigotimes_{aux} 1) R(t/t').
$$
Hence the same relation holds for the monodromy matrix of size $L$, 
$\bigotimes_{\ell =1}^L G_{p, \ell} (t)$ with $
G_{p, \ell} (t) =  G_p (t)$ at the site $\ell$. 
Therefore, the $\tau^{(2)}_p$-matrix, 
\be
\tau^{(2)}_p (t) := {\rm tr}_{aux} ( \bigotimes_{\ell=1}^L G_{p, \ell} ( \omega t)) \ \ \ \ {\rm for} \ \  t \in \CZ , 
\ele(tau2)
form a commuting
family  of $(\stackrel{L}{\otimes} \CZ^N)$-operators. The $\ZZ_N$-operators $X, Z$ of $\CZ^N$ at the site $j$ give rise to the Weyl operators $X_j Z_j$ of $\stackrel{L}{\otimes} \CZ^N$ with the relations: $Z_i X_j = \omega^{\delta_{i, j}} X_j Z_i$, $[Z_i, Z_j]= [X_i, X_j]=0$ and $Z_j^N= X_j^N=1$.  Then the spin-shift operator of $\stackrel{L}{\otimes} \CZ^N$, $X:= \prod_{j=1}^L X_j$, defines the $\ZZ_N$-charge $Q$, and commutes with $\tau^{(2)}_p (t)$.

The rapidities  of $N$-state CMP are elements in the genus $(N^3-2N^2+1)$ curve ${\goth W}$ in the projective 3-space $ \PZ^3$, defined by the equivalent sets of equations:
\begin{eqnarray}
{\goth W}  : & \left\{ \begin{array}{l}
ka^N + k'c^N = d^N , \\
kb^N + k'd^N = c^N  \end{array}  \right.   
\Longleftrightarrow  
\left\{ \begin{array}{l}
a^N + k'b^N = k d^N  , \\
k'a^N + b^N = k c^N  \end{array}  \right.  \Longleftrightarrow  
\left\{ \begin{array}{l}
k x^N  = 1 -  k'\mu^{-N}  , \\
k y^N  = 1 -  k' \mu^N  \end{array}  \right.
\label{rapidC}
\end{eqnarray}
where $[a, b, c, d] \in \PZ^3$, $(x, y, \mu)=(\frac{a}{d}, \frac{b}{c}, \frac{d}{c}) \in \CZ^3$,  $k'$ is a parameter with $k^2= 1 - k'^2 \neq 0, 1$. 
By eliminating $\mu^N$ in the last set of above equations , and using the variables $t:= xy, \lambda:= \mu^N$, one arrives the hyperelliptic curve of genus $N-1$, $ t^N = \frac{(1- k' \lambda )( 1 - k' \lambda^{-1}) }{k^2 }$, 
as a $N^2$-unramified quotient of (\ref{rapidC}). The rapidities possess a large symmetry group, in which the following two will be needed in our later discussion,
\bea(lll)
U :  [a,b,c,d] \mapsto [\omega a,b,c,d], & \Longleftrightarrow   (x, y, \mu) \mapsto ( \omega x, y, \mu), & \Longleftrightarrow (t, \lambda) \mapsto (\omega t, \lambda) ;  \\
C :  [a,b,c,d] \mapsto [b,a, d, c ], & \Longleftrightarrow (x, y, \mu) \mapsto ( y, x , \mu^{-1} ), & \Longleftrightarrow (t, \lambda) \mapsto (t, \lambda^{-1}) .
\elea(UC)

The Boltzmann weights
$W_{p,q},\overline{W}_{p,q}$ of the CPM, depending on two rapidities $p, q \in {\sf W}$, are defined by
$$
\begin{array}{ll}
\frac{W_{p,q}(n)}{W_{p,q}(0)}  = \prod_{j=1}^n
\frac{d_pb_q-a_pc_q\omega^j}{b_pd_q-c_pa_q\omega^j} , &
\frac{\overline{W}_{p,q}(n)}{\overline{W}_{p,q}(0)} 
 = \prod_{j=1}^n
\frac{\omega a_pd_q-
d_pa_q\omega^j}{ c_pb_q- b_pc_q \omega^j},
\end{array}
$$
which satisfy the $N$-periodicity property for $n$ by the constraint (\ref{rapidC}) of rapidities. 
The CPM transfer matrix of size $L$ with 
periodic boundary condition, $L+1=1$, is the $(\stackrel{L}{\otimes} \CZ^N)$-operator defined by 
\begin{eqnarray}
T_{\rm cp} (p; q)_{\sigma_1 , \ldots , \sigma_L}^{\sigma_1' , \ldots , \sigma_L'} = \prod_{l=1}^L
\overline{W}_{p,q}(\sigma_l - \sigma_l')
W_{p,q}(\sigma_l - \sigma_{l+1}') \ , \ \ \sigma_l, \sigma_l' \in
\ZZ_N .
\label{Tpq}
\end{eqnarray}
For a fixed $p \in {\goth W} $, $\{ T_{\rm cp} (p; q) \}_{q \in {\goth W}} $ form a commuting family of operators by the following well-known 
star-triangle relation of Boltzmann weights:
$$
\sum_{n=0}^{N-1} \overline{W}_{qr}(\sigma' - n) W_{pr}(\sigma - n) \overline{W}_{pq}(n - \sigma'') = R_{pqr} W_{pq}(\sigma - \sigma' )  \overline{W}_{pr}(\sigma' - \sigma'') W_{qr}(\sigma -\sigma'') ,
$$
where $R_{pqr}= \frac{f_{pq}f_{qr}}{f_{pr}}$ with $f_{pq}= \bigg( \frac{{\rm det}_N( \overline{W}_{pq}(i-j))}{\prod_{n=0}^{N-1} W_{pq}(n)}\bigg)^{\frac{1}{N}}$.
Then $T_{\rm cp} (p; q)$ commutes with $X$ and the spatial translation operator $S_R$ (which defines the total momentum $P \in \ZZ_L$). Denote $\widehat{T}_{\rm cp}(p; q) := T_{\rm cp}(p; q) S_R$. 
The transfer matrix $T_{\rm cp} (p; q)$ can be derived from $\tau^{(2)}_p (t_q)$ with $p \in {\goth W}$ as the auxiliary ``$Q$''-operator, as discussed in \cite{Bax} on the TQ-relation of 
the eight-vertex model.  
One arrives $\tau^{(2)}T_{\rm cp}$-relation ((4.20) in \cite{BBP})\footnote{$\tau^{(2)}_p ( q ), T_{\rm cp}(p; q)$ in this paper are the operators $\tau^{(2)}_{k=0, q}$, $T_q$ in \cite{BBP} respectively.} using the automorphism $U$ in (\req(UC)):
\begin{eqnarray}
&\tau^{(2)}_p(t_q) T_{\rm cp}( p; Uq)  = \varphi_p(q) T_{\rm cp}(p(; q) + \overline{\varphi}_p(Uq) X T_{\rm cp}(p; U^2q) \ \label{tauT} \\
\Longleftrightarrow & \tau^{(2)}_p(t_q) = \bigg( \varphi_p(q) T_{\rm cp}(p;q) + \overline{\varphi}_p(Uq) X T_{\rm cp}(p; U^2 q) \bigg) T_{\rm cp}(p; Uq)^{-1} \label{tau2T} 
\end{eqnarray}
where $
\varphi_p(q):= (\frac{(y_p -\omega x_q)(t_p -t_q)}{y_p^2(x_p - x_q)})^L$, $ \overline{\varphi}_p(q) := (\frac{\omega \mu_p^2(x_p -  x_q)(t_p - t_q)}{y_p^2(y_p - \omega x_q)})^L $. By (\ref{tau2T}) and the commutativity of $T_{\rm cp}(p; *)$, $
[ \tau^{(2)}_p(t_q) , T_{\rm cp}(p; q') ] = 0 $  for $ p, q, q' \in {\goth W}$. 
The fusion operators $\tau^{(j)}(t)$ for $0 \leq j \leq N$ are determined by the following $T_{\rm cp}\widehat{T}_{\rm cp}$-relation for $0 \leq j \leq N$, ((3.46) for $(l,k)=(j, 0)$ in \cite{BBP}) with $\tau^{(0)}_p:= 0 $, $\tau^{(1)}_p:= I$,
\be
T_{\rm cp}(p; q) \widehat{T}_{\rm cp}(p; CU^j q ) =  r_{p, q} h_{j ; p, q} \bigg( \tau^{(j)}_p(t_q) + \frac{z(t_q)z(\omega t_q)\cdots z(\omega^{j-1} t_q)}{\alpha_p (\lambda_q)} X^j \tau_p^{(N-j)} (\omega^j t_q) \bigg) 
\ele(TThat)
where $
z(t)= \bigg(\frac{\omega \mu_p^2 (x_p y_p - t )^2 }{y_p^4}\bigg)^L$, $
\alpha_p ( \lambda_q ) = \bigg(\frac{k'(1-\lambda_p \lambda_q)^2}{ \lambda_q (1-k' \lambda_p)^2 }\bigg)^L$ and  $r_{p, q}  = \bigg(\frac{N(x_p - x_q) (y_p - y_q) (t_p^N-t_q^N)}{(x_p^N - x_q^N) (y_p^N - y_q^N)(t_p - t_q) } \bigg)^L$ , $
h_{j ; p, q} =  \bigg( \prod_{m=1}^{j-1} \frac{y_p^2 (x_p - \omega^m x_q)}{(y_p - \omega^m x_q)(t_p - \omega^m t_q) }\bigg)^L $.  
By (\req(TThat)), one can derive the fusion relations of $\tau^{(j)}$s ((4.27) of \cite{BBP} ):
\bea(cl)
\tau^{(j)}_p(t) \tau^{(2)}_p(\omega^{j-1} t) = z( \omega^{j-1} t ) X \tau^{(j-1)}_p(t) + \tau^{(j+1)}_p(t)  , & 1 \leq j \leq N , \\
\tau^{(N+1)}_p(t):= z(t ) X \tau^{(N-1)}_p( \omega t) + u (t) I  & 
\elea(Fus)
where $u (t) := \alpha_p ( \lambda) + \alpha_p ( \lambda^{-1} )$.
Note that with $\tau^{(2)}_p(t)$ in (\req(tau2)) for $p \in \PZ^3$, the validity of fusion relation (\req(Fus)) provides a characterization of the rapidity constraint (\ref{rapidC}) for 
$p \in {\goth W}$, (Theorem 1 in \cite{R04}). Using (\req(tau2T)) and (\req(Fus)), one can express  $\tau_p^{(j)}$ in terms of $T_{\rm cp}(p; q)$, hence the $\tau^{(j)}T_{\rm cp}$-relations for $1 \leq j \leq N+1$, ((4.34) in \cite{BBP}):
\bea(ll)
\tau^{(j)}_p(q) =&  T_{\rm cp}(p; q) T_{\rm cp}(p; U^j q) \sum_{m=0}^{j-1} \bigg( \varphi_p(q)\varphi_p(Uq) \cdots \varphi_p(U^{m-1}q)  \times \\
&\overline{\varphi}_p(U^{m+1}q) \cdots \overline{\varphi}_p(U^{j-1}q) 
T_{\rm cp}(p; U^m q)^{-1} T_{\rm cp}(p; U^{m+1}q)^{-1} X^{j-m-1} \bigg) .
\elea(taujT)
By substituting (\req(taujT)) in (\req(TThat)), one obtains a single functional equation of $T_{\rm cp}(p; q)$ ((4.40) of \cite{BBP}):
\bea(ll)
 \widehat{T}_{\rm cp}(p; q) &=    \sum_{m=0}^{N-1} C_{m; p}(q) T_{\rm cp}(p; q) T_{\rm cp}(p; U^m q)^{-1} T_{\rm cp}(p; U^{m+1}q)^{-1} X^{-m-1} ,
\elea(TT)
where $C_{m; p}(q) =   \varphi_p(q) \cdots \varphi_p(U^{m-1}q) 
\overline{\varphi}_p(U^{m+1}q) \cdots \overline{\varphi}_p(U^{N-1}q) ( \frac{N y_p^{2N-2}(y_p - y_q)(y_p -  x_q) }{ (y_p^N - y_q^N)(y_p^N - x_q^N ) })^L $.

\subsection{Bethe Ansatz and Onsager algebra symmetry in superintegrable chiral Potts model}\label{subsec:OAsym}
For CPM in the superintegrable case, i.e., the rapidity $p$ given by
$\mu_p = 1$, $ x_p = y_p = \eta^{\frac{1}{2}}$, where $\eta := ( \frac{1-k'}{1+k'})^{\frac{1}{N}}$, simplification occurs for the functional relations. We shall omit the label $p$ appeared in all operators for the superintegrable case, simply write  $\tau^{(2)}(t)$, $T_{\rm cp}(q)$, {\it etc}.
As $q$ tends to $p$, to the first order of small $\varepsilon$, one has ((1.11) in \cite{AMP}),
$$
T_{\rm cp} (q) = {\bf 1} [ 1 + \varepsilon (N-1)L ] + \varepsilon H (k') 
$$
where $H(k')= H_0 + k' H_1$ is the $\ZZ_N$-quantum chain in \cite{GR, HKN}, with the expression: $H_0 = - 2\sum_{\ell=1}^L \sum_{n=1}^{N-1}
\frac{ X_\ell^n }{1-\omega^{-n}}$, $H_1 = - 2 \sum_{\ell=1}^L \sum_{n=1}^{N-1}
\frac{Z_\ell^nZ_{\ell+1}^{-n}}{1-\omega^{-n}}$, which satisfy the Dolan-Grady relation \cite{DG}, hence generate the Onsager algebra representation where only the spin-$\frac{1}{2}$ subrepresentation occurs as irreducible factors \cite{AMP, B88, B94, D91, R91}. Through the gauge transform by $ M = {\rm dia}. [1,\eta^{\frac{1}{2}}]$, the monodromy matrix $G (t)$ in (\req(G)) becomes
$$
\widetilde{G} (\widetilde{t})  = \left( \begin{array}{cc}
       1 - \widetilde{t} X & ( 1 - \omega  X )  Z   \\
       - \widetilde{t} (1 - X ) Z^{-1} & - \widetilde{t}   + \omega   X  
\end{array} \right), \ \ \widetilde{t} = \eta^{-1}t , 
$$
which is again a Yang-Baxter solution (\req(G)) for $a=b=c=d=1$. Hence $\tau^{(2)}(t)= \widetilde{\tau}^{(2)}(\widetilde{t})$, where $\widetilde{\tau}^{(2)}(\widetilde{t})$ is the trace of the $L$-size monodromy matrix associated to $G(\omega \widetilde{t})$. 
Write $\tau^{(j)}(t)= \widetilde{\tau}^{(j)}(\widetilde{t})$, the fusion relation (\req(Fus)) has the form:  
\bea(cll)
\widetilde{\tau}^{(j)}(\widetilde{t}) \widetilde{\tau}^{(2)}(\omega^{j-1} \widetilde{t}) &= (1 - \omega^{j-1} \widetilde{t} )^{2L} ~  \widetilde{\tau}^{(j-1)}(\widetilde{t}) ~ \omega^L X ~ + \widetilde{\tau}^{(j+1)}(\widetilde{t})  , & 1 \leq j \leq N \ ,  \\
\widetilde{\tau}^{(N+1)}(\widetilde{t}) & = (1 - \widetilde{t} )^{2L} ~ \widetilde{\tau}^{(N-1)}( \omega \widetilde{t}) \omega^L X  ~ + 2 (1 - \widetilde{t}^N  )^L  \ . &
\elea(Fus-) 
By examining commutators of $H_k$ with the entries of the monodromy matrix constructed from $G(\widetilde{t})$, one can show $[H_k, \tau^{(2)}(t)]=0$ for $k=0,1$. It follows the Onsager algebra symmetry of $\tau^{(j)}$-model (Theorem 1 of \cite{R05}). However, the understanding of the detailed nature of Onsager algebra symmetry in the superintegrable CPM still requires the full knowledge about eigenvalues of CPM transfer matrix, which was solved by the Bethe-ansatz method in \cite{AMP, B93, B94} as follows.

For parameters $v_1, \ldots, v_{m_p}$ with $(-v_i )^N \neq 0, 1$ and $v_i v_j^{-1} \neq 1, \omega$ for $i \neq j$, consider the rational function 
\be
{\cal P} ( \widetilde{t} )  =  \omega^{-P_b}
\sum_{j=0}^{N-1}
\frac{(1-\widetilde{t}^N)^L(\omega^j\widetilde{t})^{-P_a-P_b} }{(1-
\omega^j \widetilde{t})^L F(\omega^j \widetilde{t}) F(
\omega^{j+1} \widetilde{t})} , \ \ F(\widetilde{t}):= \prod_{i =1}^{m_p} ( 1 + \omega v_i \widetilde{t} ) 
\ele(cpP)
where $P_a, P_b$ are integers satisfying $
0 \leq r (:= P_a+P_b) \leq N-1$, $ P_b-P_a \equiv Q+L \pmod{N}$.  
${\cal P}( \widetilde{t} )$ is invariant under $\widetilde{t}
\mapsto \omega \widetilde{t}$, hence depending only on $\widetilde{t}^N$. The criterion of
 ${\cal P} ( \widetilde{t} )$ as a $\widetilde{t}$-polynomial is the following constraint for $v_j$s, ((4.4) in \cite{AMP}, (6.22) in \cite{B93}):
\be
( \frac{v_i + \omega^{-1}}{v_i+ \omega^{-2}} )^L
= - \omega^{-r} \prod_{l=1}^{m_p}
\frac{v_i - \omega^{-1}v_l}{v_i - \omega v_l} ,
\ \ \ i = 1, \ldots, m_p .
\ele(CPMBe)
Here the non-negative integer $m_p$ satisfies the relation $LP_b \equiv m_p (Q-2P_b-m_p) \pmod{N}$. The total momentum $P$ is given by $
e^{{\rm i}P}  =  \omega^{-P_b}\prod_{i =1}^{m_p} \frac{1 + \omega v_i  }{ 1 + \omega^2 v_i}$. 
The above relation is indeed the Bethe equation of $\tau^{(2)}$-model (Theorem 3 of \cite{R05}). Then ${\cal P}(\widetilde{t})$ is a simple $\widetilde{t}^N$-polynomial of degree $m_E=[\frac{(N-1)L -r-2m_p}{N}]$ with negative real roots, (Theorem 2 of \cite{R05}). Let $s_1, \ldots s_{m_E}$ be the $\widetilde{t}^N$-zeros of ${\cal P}( \widetilde{t} )$, and define 
${\cal G} (\lambda) = \prod_{j=1}^{m_E} \frac{\lambda + 1 \pm (\lambda - 1) w_j}{2\lambda}$ where $w_j :  = (\frac{ s_j - \eta^{-2N}}{  s_j -1 })^{\frac{1}{2}}$. 
Then $
\frac{{\cal P}(\widetilde{t})}{{\cal P}(1)} = {\cal G} (\lambda){\cal G} (\lambda^{-1})$. 
One has the following expression of $T_{\rm cp}(q)$-eigenvalues ((1.11) in \cite{AMP} and (21) in \cite{B94}):  
\be
T_{\rm cp}(q) = N^L \frac{(\eta^{\frac{-1}{2}} x_q-1)^L}{(\eta^{\frac{-N}{2}} x_q^N-1)^L} (\eta^{\frac{-1}{2}}x_q)^{P_a}(\eta^{\frac{-1}{2}}y_q)^{P_b}\mu_q^{-P_\mu} \frac{F(\widetilde{t}_q) }{F(1)} {\cal G} (\lambda_q) \ (= e^{-{\rm i}P} \widehat{T}_{\rm cp}(q)) ,
\ele(Tqval)
which gives the energy value of $H(k')$ ((2.23) of \cite{AMP}):
$$
E = 2 P_\mu + N m_E - (N-1)L + k' \bigg( (N-1)L -2P_\mu-Nm_E + 2(P_b-P_a)\bigg) + N(1-k') \sum_{j=1}^{m_E} \pm w_j .
$$
Therefore the $\tau^{(2)}$-degenerate states associated to the Bethe roots $v_i$s form an irreducible Onsager-algebra-representation space of dimension $2^{m_E}$, which we associate the following normalized CPM transfer matrix\footnote{The $Q$-operator here differs from the ${\sc Q}_{cp}$ in \cite{R05} by a scale factor: $(\eta^{\frac{-1}{2}}x_q)^{P_a}(\eta^{\frac{-1}{2}}y_q)^{P_b}\mu_q^{-P_\mu} Q(q) = {\sc Q}_{cp}(q)$.}:
$$
Q (q) = \frac{T_{\rm cp} (q)(1- \eta^{\frac{-N}{2}} x_q^N)^L}{ N^L (1 - \eta^{\frac{-1}{2}} x_q )^L(\eta^{\frac{-1}{2}}x_q)^{P_a}(\eta^{\frac{-1}{2}}y_q)^{P_b}\mu_q^{-P_\mu}} \ ( =: e^{-{\rm i}P} \widehat{Q} (q) ).
$$
The $Q$-eigenvalues and the functional equation (\req(TT)) now become
\begin{eqnarray}
&Q (q) =  \frac{F(\widetilde{t}_q) }{F(1)} {\cal G} (\lambda_q) , \ \ \  \ \widehat{Q}( C q ) = e^{{\rm i}P} \frac{F(\widetilde{t}_q) }{F(1)} {\cal G} (\lambda_q^{-1} ) , \label{Qval} \\
&\widehat{Q}( C q ) =  \frac{\omega^{-P_b}}{N^L }   \sum_{m=0}^{N-1} \frac{(1- \widetilde{t}_q^N)^L \omega^{-mr} }{(1- \omega^m \widetilde{t}_q)^L}  Q(q) Q(U^m q)^{-1}Q(U^{m+1}q)^{-1} \ .
\label{QS}
\end{eqnarray}
The relations, (\req(tauT)) (\req(taujT)) (\req(TThat)), now  become
\begin{eqnarray}
&\widetilde{\tau}^{(2)}(\widetilde{t}_q) Q (Uq)  = \omega^{-P_a}(1 -\widetilde{t}_q)^L Q (q) + \omega^{P_b} (1 -  \omega \widetilde{t}_q)^L  Q (U^2 q)  ,
\label{tauT-} \\
&\widetilde{\tau}^{(j)}(\widetilde{t}_q) 
=  \omega^{(j-1)P_b } \sum_{m=0}^{j-1} \bigg( \frac{\prod_{ k=0}^{j-1} (1- \omega^k \widetilde{t}_q)^L}{(1- \omega^m \widetilde{t}_q)^L} Q (q)
Q (U^m q)^{-1} Q(U^j q) Q(U^{m+1}q)^{-1} \omega^{-r} \bigg)  ,
\label{taujT-} \\
&N^L \widetilde{t}_q^r Q(q) \widehat{Q}( C U^j q )= \frac{\omega^{-jP_b}(1-  \widetilde{t}_q^N)^L}{\prod_{m=0}^{j-1} (1 - \omega^m \widetilde{t}_q)^L } \widetilde{\tau}^{(j)}(\widetilde{t}_q)   + \frac{ \omega^{-jP_a}(1-  \widetilde{t}_q^N)^L }{\prod_{m=j}^{N-1} (1 - \omega^m \widetilde{t}_q)^L} \widetilde{\tau}^{(N-j)} (\omega^j \widetilde{t}_q), \ 0 \leq j \leq N.  \label{TT-}
\end{eqnarray}
By (\ref{Qval}), the relation (\ref{tauT-}) yield ((6.18) in \cite{B93}): 
\be
\widetilde{\tau}^{(2)}(\widetilde{t}_q)  F(\omega \widetilde{t}_q)  = \omega^{-P_a} (1 - \widetilde{t}_q)^L   F( \widetilde{t}_q)   + \omega^{P_b} (1 - \omega \widetilde{t}_q)^L     F(\omega^2 \widetilde{t}_q).
\ele(CPt2)
Using (\ref{taujT-}) and (\ref{TT-}), one obtains the $\widetilde{\tau}^{(j)}$-polynomial expression (Theorem 3 $(ii)$ of \cite{R05}):
\bea(ll)
\widetilde{\tau}^{(j)}(\widetilde{t}) =  \omega^{(j-1)P_b} \prod_{ k=0}^{j-1} (1- \omega^k \widetilde{t})^L  \sum_{m=0}^{j-1}  \frac{ F(\widetilde{t}) F(\omega^j \widetilde{t}) \omega^{-m ( P_a +P_b)}}{(1- \omega^m \widetilde{t})^L F(\omega^m \widetilde{t})F(\omega^{m+1} \widetilde{t})}, &  2 \leq j \leq N , \\
\widetilde{t}^r F(\widetilde{t}) F(\omega^j \widetilde{t}) {\cal P} (\widetilde{t})  = \frac{\omega^{-jP_b}(1-  \widetilde{t}^N)^L}{\prod_{m=0}^{j-1} (1 - \omega^m \widetilde{t})^L } \widetilde{\tau}^{(j)}(\widetilde{t})   +  \frac{\omega^{-jP_a} (1-  \widetilde{t}^N)^L  }{\prod_{m=j}^{N-1} (1 - \omega^m \widetilde{t})^L} \widetilde{\tau}^{(N-j)} (\omega^j \widetilde{t})  , & 0 \leq j \leq N .
\elea(CPtj)

\section{Six-vertex Model at Roots of Unity}
\subsection{Bethe equation of the six-vertex model}\label{subsec:Bethe6}
The transfer matrix of the six-vertex model of an {\it even} size $L$ is the $(\stackrel{L}{\otimes} \CZ^2)$-operator constructed from the Yang-Baxter solution\footnote{Here we use the convention in \cite{FM01}, Eq. (1.3): $a={\rm i} \sinh \frac{1}{2}(v- {\rm i} \gamma)$, $b=- {\rm i} \sinh \frac{1}{2}(v + {\rm i} \gamma)$, $c= - {\rm i} \sinh {\rm i} \gamma$, with the variables $z= -e^v , {\sf q}= -e^{{\rm i} \gamma}$. Note that the ${\sf q}$ here differs from the $q$ in \cite{FM01} by minus sign for its connection with $U_{\sf q}(\widehat{sl_2})$ as  Eq. (2.3) in \cite{De05}, where $-e^{-z}$ is equal to $z^{\frac{1}{2}}$ here. }
$$
L = \left( \begin{array}{cc}
    z^{\frac{1}{2}} {\sf q}^{\frac{-\sigma^{\rm Z} }{2}} - z^{\frac{-1}{2}} {\sf q}^{\frac{\sigma^{\rm Z} }{2}}, & ({\sf q}-{\sf q}^{-1}) \sigma_- \\
({\sf q}-{\sf q}^{-1}) \sigma_+ , & z^{\frac{1}{2}} {\sf q}^{\frac{\sigma^{\rm Z} }{2}} - z^{\frac{-1}{2}} {\sf q}^{\frac{- \sigma^{\rm Z} }{2}}
\end{array} \right) \ , \ \ \ \sigma^{\rm Z}, \sigma_{\pm}: {\rm Pauli \ \ matrix}, 
$$
for the R-matrix
$$
R (z) = \left( \begin{array}{cccc}
        z^{\frac{-1}{2}} {\sf q} - z^{\frac{1}{2}} {\sf q}^{-1}  & 0 & 0 & 0 \\
        0 &z^{\frac{-1}{2}} - z^{\frac{1}{2}} & {\sf q}-{\sf q}^{-1} &  0 \\ 
        0 & {\sf q}-{\sf q}^{-1} &z^{\frac{-1}{2}} - z^{\frac{1}{2}} & 0 \\
     0 & 0 &0 & z^{\frac{-1}{2}} {\sf q} - z^{\frac{1}{2}} {\sf q}^{-1} 
\end{array} \right),
$$
as the trace of monodromy matrix: $T (z) = {\rm tr}_{aux} ( \bigotimes_{\ell=1}^L L_\ell (z))$ for $z \in \CZ$. The logarithmic $z\frac{d}{dz}$-derivative of $T (z)$ at $z= {\sf q}^{-1}$ gives rise to the XXZ chain with the periodic boundary condition:
$$
H_{XXZ} = -\frac{1}{2} \sum_{\ell =1}^L ( \sigma_\ell^1 \sigma_{\ell+1}^1 +
\sigma_\ell^2 \sigma_{\ell+1}^2 + \triangle \sigma_\ell^3 \sigma_{\ell+1}^3 ) , \ \ \ \ \ \triangle = \frac{1}{2}({\sf q}+{\sf q}^{-1}),
$$
which has been a well-studied Hamiltonian initiated by Bethe in 1931 \cite{Be}. A major result of the study is that the ground state energy for the value $S^Z (= \frac{1}{2} \sum_\ell \sigma_\ell^Z)$ is determined by an appropriate solution of the following Bethe equation for the variable $v = - z^{-1}$:
\be
(\frac{  v_i + {\sf q}^{-1}}{ v_i +{\sf q} })^L =  - {\sf q}^{-L+2m} \prod_{ l =1 }^m \frac{  v_i  - {\sf q}^{-2} v_l  }{  v_i   - {\sf q}^2 v_l } , \ \ \ \ \ m = \frac{L}{2} - |S^Z| .
\ele(BeXXZ)
The Bethe-equation technique was further extended to the method of Baxter's TQ-relation in eight-vertex model; when applying to the six-vertex model, there exists a non-degenerated commuting family of $Q$-operators with the following relation (see, Chapter 9 of \cite{Bax}):
\be
T (z)  Q (s) = {\sf q}^{-2|S^Z|} (1 - {\sf q}z) ^L Q (U^{-1}s) +  (1 - {\sf q}^{-1}z)^L Q (U s) .  
\ele(TQ6)
Here $s$ is a suitable multi-valued complex coordinate of $z$, and $U$ is a $s$-automorphism which induces the $z$-transform: $z \mapsto {\sf q}^2 z$. Note that there are many such $Q$-operators, however all give the same Bethe equation (\req(BeXXZ)) through Eq.(\req(TQ6)).

\subsection{Evaluation polynomial and fusion relation of six-vertex model at roots of unity}\label{subsec:Pol6V}
For the root of unity case with ${\rm q}^{2N}=1$, i.e. ${\rm q}^2= \omega$,  the six-vertex model possesses the $sl_2$-loop algebra symmetry \cite{DFM}. The Bethe state corresponding to the Bethe roots  is the ``highest weight'' vector of an irreducible representation of the $sl_2$-loop algebra, with the evaluation parameters characterized by the Drinfeld polynomial \cite{De05, FM01}. By studying the creation $sl_2$-loop current operator in the ABCD-algebra, the Drinfeld polynomial for a Bethe root $\{ v_i \}_{i=1}^m$ of  (\req(BeXXZ)) is given by Eq. (3.9) in \cite{FM01}. Denote $\widetilde{t}= {\sf q}z$, and define the integer $r$ by  $r \equiv \frac{L}{2}- m \pmod{N}$, $0 \leq r \leq N-1$. The Drinfeld polynomial is indeed the $\widetilde{t}^N$-polynomial associated the following polynomial $P(\widetilde{t})$\footnote{In \cite{FM01}, the evaluation function is given by (3.9) there: $Y(v) = \sum_{j = 0}^{N-1} \overline{a}(v +2 (j+1) {\rm i}\gamma  )$, where $\overline{a}(v)=   \frac{\sinh^L \frac{1}{2}(v-{\rm i}\gamma)}{\prod_{i =1}^m \sinh \frac{1}{2}(v_i -v) \sinh \frac{1}{2}(v_i - v + 2 {\rm i} \gamma)}$ in (2.47) of \cite{FM01}. In terms of variables $z, \widetilde{t}$ and Bethe roots $v_i$s here, $\overline{a}(v)
= 2^{2m-L} (\prod_{i=1}^m v_i)({\sf q}^{-1}z)^{\frac{-L}{2}+m}  \frac{(1- {\sf q}^{-1}z)^L}{\prod_{i=1}^m (1+v_i z )(1+ \omega^{-1} v_i z)}$, which implies $Y(v) = 2^{2m-L} (\prod_{i=1}^m v_i )  \widetilde{t}^{\frac{-L}{2}+m +r} P(\widetilde{t})$.},
\be
P(\widetilde{t})=  \sum_{j=0}^{N-1} \frac{(1- \omega^j \widetilde{t})^L (\omega^j \widetilde{t})^{ -r}}{F(\omega^j \widetilde{t}) F(\omega^{j+1} \widetilde{t})} \ , \ \ F(\widetilde{t}) : = \prod_{i =1}^{m} ( 1 + {\sf q}^{-1} v_i \widetilde{t}  ),
\ele(6vP)
which has a similar form as (\req(cpP)). Indeed with $F(\widetilde{t})$ in (\req(cpP)) or (\req(6vP)), let $H(\widetilde{t}) = \frac{1 - \widetilde{t}^N}{1-\widetilde{t}}, 1-\widetilde{t}$ in CPM, six-vertex model respectively. The function $P(\widetilde{t})$ defined by $
P(\widetilde{t})=  \sum_{j=0}^{N-1} \frac{H ( \omega^j \widetilde{t})^L (\omega^j \widetilde{t})^{ -r}}{F(\omega^j \widetilde{t}) F(\omega^{j+1} \widetilde{t})}$ is invariant under $\widetilde{t} \mapsto \omega \widetilde{t}$. The condition on roots of $F(\widetilde{t})$ so that $P(\widetilde{t})$ is a polynomial is provided by Bethe equation (\req(CPMBe)), (\req(BeXXZ)) respectively. Define the $T^{(2)}$-operator by $T^{(2)}(\widetilde{t}) = z^{\frac{L}{2}} T(z)$ in the six-vertex model, and  
$\frac{\omega^{-P_b} (1- \widetilde{t}^N)^L \widetilde{\tau}^{(2)}(\omega^{-1} \widetilde{t}) }{(1- \omega^{-1} \widetilde{t})^L(1-  \widetilde{t})^L}$ in CPM. Then  Eqs. (\ref{tauT-}), (\req(TQ6)) are combined into one $T^{(2)}Q$-relation: 
\be
T^{(2)}( \widetilde{t})  Q (q) = \omega^{-r} H(\widetilde{t})^L Q (U^{-1}q) +  H ( \omega^{-1} \widetilde{t})^L Q (Uq), \ \ U^N = 1 .  
\ele(T2Q)
The $T^{(j)}$-operators are defined recursively through the following fusion relation for $j \geq 1$ by setting $T^{(0)}= 0$, $T^{(1)} = H( \omega^{-1}\widetilde{t})^L $, 
\begin{eqnarray}
T^{(j)}( \widetilde{t}) T^{(2)}(\omega^{j-1} \widetilde{t}) = \omega^{-r} H(\omega^{j-1} \widetilde{t})^L T^{(j-1)}( \widetilde{t})  + H(\omega^{j-2}\widetilde{t})^L T^{(j+1)}( \widetilde{t}). \label{Tfus}  
\end{eqnarray}
By (\req(T2Q)), the induction-argument yields the $T^{(j)}Q$-relation for $j \geq 0 $:
\be
T^{(j)}( \widetilde{t})
= Q(U^{-1} q) Q(U^{j-1} q)  \sum_{k =0}^{j-1}( \omega^{-k r} H( \omega^{k -1} \widetilde{t})^L  
Q(U^{k -1} q)^{-1}  Q(U^k q)^{-1}). 
\ele(TjQ)
By this, one obtains the boundary condition of the fusion relation:
\begin{eqnarray}
T^{(N+1)}(\widetilde{t})  = \omega^{-r} T^{(N-1)}( \omega \widetilde{t})
  + 2 H ( \omega^{-1} \widetilde{t})^L . \label{TfBdy}
\end{eqnarray}
In CPM case, with the identification $T^{(j)}( \widetilde{t}) = \frac{\omega^{-(j-1)P_b} (1- \widetilde{t}^N)^L \widetilde{\tau}^{(j)}(\omega^{-1}\widetilde{t}) }{\prod_{k=-1}^{j-2}(1- \omega^k \widetilde{t})^L}$, Eqs. (\ref{Tfus})-(\ref{TfBdy}) are the same as Eqs. (\req(Fus-)), (\req(taujT-)). While in six-vertex model, the fusion relation and $T^{(j)}Q$-relation hold for any $Q$-operator satisfying $T^{(2)}Q$-relation (\req(T2Q)). 

For a polynomial $F(\widetilde{t})$ with Bethe roots $v_i$s, by (\req(TjQ)) the corresponding $T^{(2)}$-eigenvalue is determined by the relation
\be
T^{(2)}( \widetilde{t})  F(\widetilde{t}) = \omega^{-r} H(\widetilde{t})^L F( \omega^{-1}\widetilde{t})  +  H ( \omega^{-1} \widetilde{t})^L F( \omega \widetilde{t}).   
\ele(TF)
Using Eq.(\ref{Tfus}) to express $T^{(j+1)}$ in $T^{(j)}$ and $T^{(j-1)}$, by induction argument, one obtains the form of $T^{(j)}$-eigenvalues from Eq.(\req(TF)): 
\be
T^{(j)}(\widetilde{t}) = F(\omega^{-1} \widetilde{t}) F( \omega^{j-1}\widetilde{t}) \sum_{k=0}^{j-1} \frac{ H(\omega^{k-1}\widetilde{t})^L \omega^{-kr} }{F(\omega^{k-1}\widetilde{t})F(\omega^k \widetilde{t})} \ , \ \ \ \ j \geq 1 ,
\ele(TjF)
which implies 
\be
\widetilde{t}^r F( \omega^{-1} \widetilde{t}) F( \omega^{j-1}\widetilde{t}) P(\widetilde{t}) = T^{(j)}(\widetilde{t}) + \omega^{-jr} T^{(N-j)}(\omega^j\widetilde{t})  , \ \ 0 \leq j \leq N.
\ele(TTj)
Eqs.(\req(TF))-(\req(TTj)) in CPM case  are the same as Eqs. (\req(CPt2)), (\req(CPtj)). Note that Eq.(\req(TTj)) is a consequence of the $QQ$-relation (\ref{TT-}) in CPM case, which encodes the detailed nature of Onsager algebra symmetry in the derivation of Eq.(\ref{Qval}). However in the case of roots-of-unity six-vertex model, the $QQ$-relation has yet been found, even though the $sl_2$-loop algebra symmetry together with evaluation parameters has already been known \cite{DFM, De05, FM01}. Based on the understanding in the CPM case, we now describe a similar, but speculated, structure about the $QQ$-relation in six-vertex case as follows. Consider the curve $W$ and its symmetries $U, C$:
$$
W: w^2 = \widetilde{t}^N , \ \ \ U:(w, \widetilde{t}) \mapsto (w, \omega \widetilde{t}) , \ \ C:(w, \widetilde{t}) \mapsto (-w,  \widetilde{t}).
$$
For odd $N$, the above curve is parametrized by $s=\widetilde{t}^{\frac{1}{2}}$, and the automorphism $\varphi: s \mapsto {\rm q} s$ gives rise to the above symmetries by $U = \varphi^{-2[\frac{N}{2}]}$, $C= \varphi^N$. The polynomial $P(\widetilde{t})$ in (\req(6vP)) are expected\footnote{The statement is true in the case $r=0$, which has been justified in \cite{De05}.} to have the simple $\widetilde{t}^N$-roots $s_1, ..., s_M$ with $P(0) \neq 0$. Define $G( w) = \prod_{j=1}^M (\sqrt{s_j}- w)$, then $\frac{G(w)G(-w)}{G(0)^2} = \frac{P( \widetilde{t})}{P(0)}$. In the eigen-space of $T(z)$ corresponding to $F(\widetilde{t})$ determined by a Bethe root, The $Q$-operator has the following $Q$-eigenvalues:
$$
Q(q) = F ( \widetilde{t}) \frac{G(w)}{G(0)}, \ \ \ \ Q(Cq) = F ( \widetilde{t}) \frac{G(-w)}{G(0)} \ \ \ \ {\rm for } \ q=(w, \widetilde{t}).
$$ 
The above conditions reveal the $sl_2$-loop symmetry of six-vertex model, as the role of Eq.(\ref{Qval}) for the Onsager algebra symmetry in CPM. Hence such a $Q$-matrix, if exits, must possess certain constraints in order to incorporate the symmetry of six-vertex model as discussed in \cite{Bax} Secs. 9.1-9.5.

\section{Discussion}
In this paper we have examined the symmetry structure of the superintegrable CPM and the six-vertex model at roots of unity by the method of functional relations. For the superintegrable CPM, exact results about the Onsager algebra symmetry of the $\tau^{(j)}$-models are obtained using the explicit form of eigenvalues of the CPM-transfer matrix. Based on common features related to evaluation parameters of the symmetry algebra representation, we discussed the Bethe ansatz of both theories in a unified manner. By this, in the six-vertex model at roots of unity, we obtained the fusion relation of  $T^{(j)}$-matrices, $T^{(j)}Q$-relation from the TQ-relation, and further indicate the special nature of $Q$-operator in accord with the required $sl_2$-loop algebra symmetry of the six-vertex model.  

\section*{Acknowledgements}
The author thanks the conference organizers for the invitation, and the local committee at Nankai University for the hospitality and hard work. I also wish to thank Professor F. Hirzebruch for the warm hospitality in November, 2005 during the author's one-month stay at Max-Planck-Institut f\"{u}r Mathematik in Bonn, where this report was completed.

\end{document}